\documentclass[sigconf,natbib]{acmart}
\usepackage{multirow}
\usepackage{amsmath}
\usepackage{tikz}

\AtBeginDocument{%
  \providecommand\BibTeX{{%
    \normalfont B\kern-0.5em{\scshape i\kern-0.25em b}\kern-0.8em\TeX}}}

\setcopyright{none}
\renewcommand\footnotetextcopyrightpermission[1]{} 
\acmConference[The 32nd ACM CIKM Conference]{Make sure to enter the correct
  conference title from your rights confirmation email}{21-25 October 2023}{UK}
\acmPrice{15.00}
\acmISBN{978-1-4503-XXXX-X/18/06}

\begin{document}

\title{Cross-Element Combinatorial Selection for Multi-Element Creative in Display Advertising}


\author{Wei Zhang, Ping Zhang, Jian Dong, Yongkang Wang,}
\author{Pengye Zhang, Bo Zhang, Xingxing Wang, Dong Wang}
\email{{zhangwei180, zhangping18, dongjian03, wangyongkang03}@meituan.com}
\email{{zhangpengye, zhangbo126, wangxingxing04, wangdong07}@meituan.com}
\affiliation{%
  \institution{Meituan}
  \city{Beijing}
  \country{China}
}


\renewcommand{\shortauthors}{Wei Zhang, et al.}

\begin{abstract}
The effectiveness of ad creatives is greatly influenced by their visual appearance. Advertising platforms can generate ad creatives with different appearances by combining creative elements provided by advertisers. However, with the increasing number of ad creative elements, it becomes challenging to select a suitable combination from the countless possibilities. The industry's mainstream approach is to select individual creative elements independently, which often overlooks the importance of interaction between creative elements during the modeling process. In response, this paper proposes a Cross-Element Combinatorial Selection framework for multiple creative elements, termed CECS. In the encoder process, a cross-element interaction is adopted to dynamically adjust the expression of a single creative element based on the current candidate creatives. In the decoder process, the creative combination problem is transformed into a cascade selection problem of multiple creative elements. A pointer mechanism with a cascade design is used to model the associations among candidates. Comprehensive experiments on real-world datasets show that CECS achieved the SOTA score on offline metrics. Moreover, the CECS algorithm has been deployed in our industrial application, resulting in a significant 6.02\% CTR and 10.37\% GMV lift, which is beneficial to the business.

\end{abstract}

\ccsdesc[500]{Information systems~Display advertising; Recommender systems; Ad Creative}

\keywords{Ad Creative, Creative Selection, Creative Interaction}


\maketitle


\section{INTRODUCTION}

In recent years, we have witnessed a growing business in online display advertising and its importance for Internet service providers. Image ads are the most widely used format since they are more compact, intuitive, and comprehensible. In the image of the advertisement, aka an $ad\: creative$, there may be background pictures, item descriptions, promotional texts, and other creative elements. And different combinations can quite vary in Click-Through Rate(CTR)\cite{azimi2012impact, mo2015image}. Therefore, it is of great value to platforms to study how to improve the visual effect of creatives.

With the increase of creative elements, it has become an emerging trend to investigate how to combine creative elements to make a better appearance and thus maximize the effectiveness of ads\cite{chen2019understanding}. However, the difficulty lies in the exponential growth in the number of candidate combinations. For example, a creative that contains three elements would involve $N_1*N_2*N_3$ possible elements combinations. More generally, assuming that there are $M$ types of creative elements, each with an average of $N$ candidates, the complexity of the problem is $N^M$. Moreover, as new creative elements are added, the number of possible combinations will increase rapidly, making it difficult to apply complex models to creative selection tasks. 

Traditionally, to deal with online performance issues caused by the excessive number of creative combinations, the creative selection is performed offline\cite{choi2010using}, thereby reducing the number of online candidates. However, the offline selection of creatives lacks the awareness of online request-level data. Therefore, some online strategies are applied, such as popularity, and preferences\cite{tang2013automatic}. However, this method requires time to accumulate online feedback. Meanwhile, some studies have attempted to classify creative elements into different types and estimate using multi-task click-through-rate(CTR) models, and then combine the elements that are estimated by different task towers\cite{richardson2007predicting}. This method solves the problem of the combinatorial explosion. The complexity here is reduced from $N^M$ to $N * M$. But for these multi-task learning routes, the cross-category interactions are not fully captured.

To this end, we propose a \textbf{C}ross-\textbf{E}lement \textbf{C}ombinatorial \textbf{S}election framework for multiple creative elements, named CECS. To capture the association between different types of creative elements, a cross-element interaction method is applied to adjust the expression of every single creative element. We further designed a cascade element selection to transform the creative combination problem into a cascade selection problem. Experimental results show that the method has a significant improvement compared with the baselines, indicating that the method can select better combinations of candidate elements. Through online real-world experiments, it is demonstrated that the method can improve the CTR and GMV, which means the selected advertisements are appealing to the users.

To summarize, the contribution of this paper can be described as follows:

\begin{itemize}
  \item We propose a cross-element interaction to tackle the relationship between different creative elements. Since the modal and information may vary, we adopt multiple attention mechanisms between each pair of creative types.
  \item We transform the creative combination problem into a cascade selection problem, the creative elements information is passed through the cascade design, so that the association in ad creatives can be integrated. 
  \item We evaluate the proposed model on real-world datasets. The offline statistics and the online feedback indicate the strong practical value of the algorithm in the industry.
\end{itemize}

\section{METHODOLOGY}

We first define the concepts and issues involved in this paper, 
then give corresponding explanations for the details of the 
different modules in the architecture.

\subsection{Problem Setup}

 \textit{\textbf{Definition 1. Creative Optimization.}} Given a user and the 
 candidate creative elements, the goal of creative optimization is to 
 find the best combination of creative elements, so as to maximize the probability of being clicked/satisfied by the user, denoted as follows:
\begin{gather}
  maximize \:  \text{ } P(A,label=1|C,u;\Theta)
\end{gather}
where $A$ means an ad creative combination and $C$ is the set of candidate elements, $label=1$ indicates the positive feedback from the user $u$, e.g. click behavior in advertising, and $\Theta$ is the model parameter to be optimized.


In real industrial scenarios, the creative elements in an ad may have multiple categories, e.g., banners, description texts, shop names, etc. Every element may contribute to the click behavior, so a design to fully explore the relationships is necessary.

\begin{figure}[htb]
  \centering
  \includegraphics[width=\linewidth]{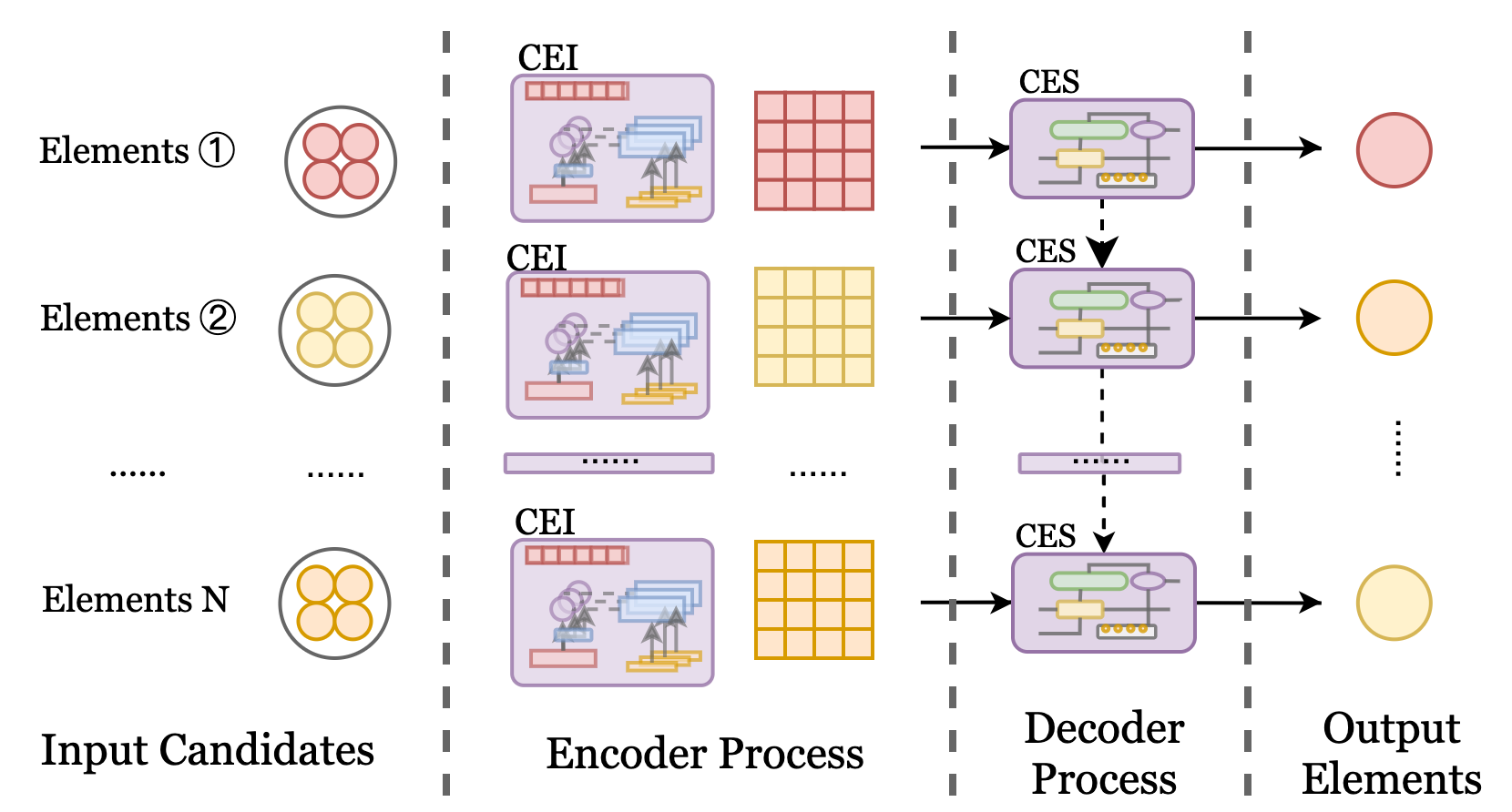}
  \Description{fig.2}
  \caption{An example of the creative selection problem with N types of creatives. The CEI and CES components are applied as the encoder process and the decoder process to capture the relationships.
  }
  \label{fig:CECS}
\end{figure}


\subsection{Model Overview}


In this section, we provide an overview of the model. The overall structure of CECS is shown in Figure \ref{fig:CECS}, which follows an encoder-decoder process design. In the encoder process, the model processes candidate creative elements in a Cross-Element Interaction (CEI), with the goal of obtaining creative elements that contain rich interaction information. In the decoder process, creative elements are sequentially estimated using a Cascade-Element Selection (CES), resulting in a combined ad.

 In this way, it is no longer necessary to enumerate every creative combination($N^M$ solutions), but only to estimate the creative elements sequentially through the model($N*M$ solutions). The inner relationship between categories is expected to be captured by the model, which will be discussed later.


\subsection{Cross-Element Interaction}
\label{sec:3.3}
In an advertisement, every creative element contributes to the final click behavior of a user, since different types of elements may be related to each other at the semantic or vision level. Traditionally, the elements are modeled separately, and the relationships that lie between different creative elements are often overlooked, which is the key point this chapter addresses.

\begin{figure}[h]
  \centering
  \includegraphics[width=0.5\linewidth]{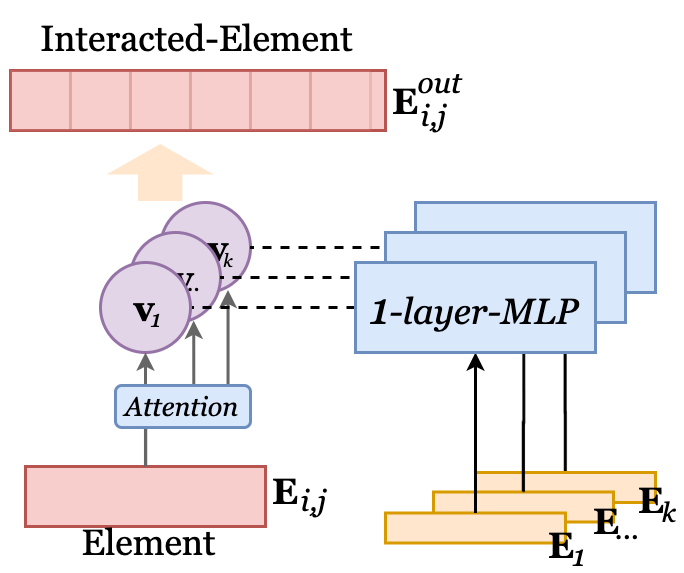}
  \caption{An illustration of procession for a single creative element in the CEI component. The result representation integrates information from all creative categories.}
  \label{fig:CEIM}
\end{figure}

In order to better integrate the relationships, we apply a simple but efficient way to combine information from all categories, so that the cross-category relationships can be captured. Inspired by the design of the parameter personalized network\cite{chang2023pepnet}, we adopt a similar interaction way, as is shown in Figure \ref{fig:CEIM}. 

Firstly, we compute the central representation of each category of candidate creative elements, which is shown as follows:
\begin{equation}
    \begin{aligned}
\boldsymbol{E}_{i} =\frac{1}{N_i} \sum_{j=1}^{N_i} \boldsymbol{E}_{i,j} \\
    \end{aligned}
\end{equation}
where $E_{i,j}$ represents the embedding of the $j$-th element of the $i$-th creative category, $N_i$ means the number of items of the $i$-th creative category. For short, the central representation is the mean pooling among the specific category. Later, we adopt a $1-layer-MLP$ to transform its semantic, i.e., $\textbf{v}_i=MLP(\textbf{E}_i)$, and the dimension is made consistent with the original representation.

Then, for each candidate's creative element $\textbf{E}_{i,j}$, we use the attention method to calculate the relationship with every transformed central representation $\textbf{v}_i$. Here the attention mechanism is applied to weight the importance of the different categories of creative elements, as is shown:

\begin{equation}
    \begin{aligned}
\textbf{E}_{i,j}^{out} = \sum_{i=1}^k \text{softmax}(\frac{\textbf{v}_{i}^T \textbf{E}_{i,j}}{\sqrt{d_{dim}}})\textbf{E}_{i,j} \\
    \end{aligned}
\end{equation}
where $k$ is the total number of creative category, $d_{dim}$ is the dimension of the embedding, $\textbf{E}_{i,j}$ is a creative element of $j$-th category, $\textbf{v}_i$ is the central representation of the $i$-th creative category, and $\textbf{E}_{i,j}^{out}$ is the output representation of the creative.

In this way, it is possible to get the encoded creatives containing interacted features by explicitly integrating the attention mechanism.


\subsection{Cascade-Element Selection}

The prediction process of creative optimization has two 
characteristics: \textbf{1)} the output is derived from the input, i.e., the predicted creative elements are derived from the candidate ones. \textbf{2)} There are relations among
the estimated creative elements, which makes it necessary to estimate in a sequential manner. The pointer network mechanism \cite{vinyals2015pointer} is a method of selecting from existing candidate solutions, and we adopt and extend it here. 

\begin{figure}[h]
  \centering
  \includegraphics[width=\linewidth]{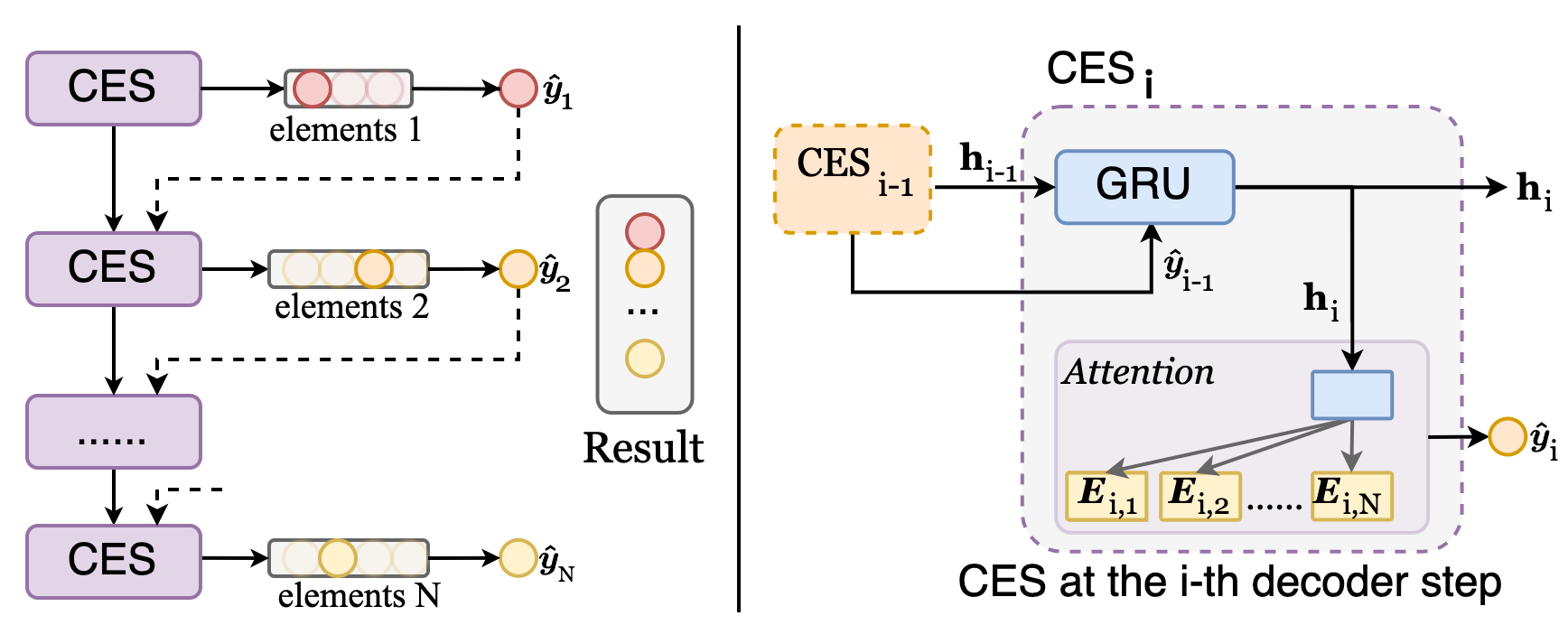}
  \Description{fig.6}
  \caption{A single step of the cascade-element selection. The creative information is transferred through the hidden vector along the GRU cells.}
  \label{fig:6}
\end{figure}

The module Cascade-Element Selection(CES for short) estimates creative elements step-by-step, so a cascade structure is applied here, as shown in Figure \ref{fig:6}. One of the key points is that in the $i$-th prediction process, the prediction status of $(i-1)$-th needs to be considered. Therefore, The Gate Recurrent Unit\cite{chung2014empirical}(GRU for short) structure is included in the CES. The hidden vector of GRU(also called the state vector in some scenarios) named $h_i$ represents the predicted state of $(i-i)$-th elements that will be transferred to the next CES unit, so the inter-relationship between output elements can be integrated. 

In the $i$-th step of the decoder, i.e., predicting the creative element of the $i$-th category, the hidden vector $\textbf{h}_{i-1}$ and $\hat y_{i-1}$ in the previous CES are fed into the current CES, as is shown in eq\eqref{eq:grufixed}:
\begin{gather}
    \begin{aligned}
        \textbf{h}_i &= GRU(\hat y_{i-1}, \textbf{h}_{i-1}) \\
        \hat y_{i} &= \sum _{j=1}^{N_i} softmax(\frac{\textbf{h}_i^T \textbf{E}_{i,j}}{\sqrt{d_{dim}}}) \textbf{E}_{i,j}
    \end{aligned}
    \label{eq:grufixed}
\end{gather}
where $\textbf{h}_i$ is the output of the GRU unit function annotated as $GRU$, $\hat y_{i-1}$ and $h_{i-1}$ are the previous output and state respectively, $N_i$ is the number of the $i$-th candidates, $\textbf{E}_{i,j}$ is the interacted representation of the candidate creative elements of the $i$-th category. For the output vector of the GRU, an attention mechanism is applied to calculate the probability among the encoded creative elements. Then we get the probability distribution over the candidates. 


 
Finally, the index of the output creative element $\hat{y}_i$ among candidates is used. Pick up the corresponding index in the candidate's creative elements, and the solution of the $i$-th creative element can be obtained.


The above process repeats $k$ times to predict every creative element, and the creative information is cascaded to the next pointer until the final creative element is predicted. Particularly, the encoded representation of the total candidates is used for the first CES unit, which acts as the overall upstream for the decoder. In this way, the cascade-element selection models the interactive relationship between different types of creative elements.

\subsection{Model Loss and Optimization}

\label{sec:3.5.1}
The estimated creative elements are derived 
from the input, so we model the output probability of each 
cascade-element selection. Therefore, we use cross-entropy to model 
the loss of one type of creative element, as is shown below:
\begin{gather}
  \begin{aligned}
  Loss_i = - y_j \log(\hat{y}_j)
  \end{aligned}
\end{gather}
where $Loss_i$ is the loss for the $i$-th category, $y_i$ and $\hat y_i$ are the ground truth and predicted result for the $i$-th category respectively.

The different categories of creative elements may result in different promotional and visual effects, so the loss of different creative elements may have a large variance. Therefore, we adopt the idea of 
uncertainty modeling\cite{kendall2018multi} to provide adaptive weight coefficients for the selection
 process of different types of creative elements. In this way, the human parameter tuning process can be reduced and better-weighted loss can be learned, shown as follows:
\begin{gather}
  Loss_{total} =  \sum_{i=1}^k \left ( \frac {1}{\sigma_i^2} Loss_i + \log {\sigma_i^2} \right )
\end{gather}
where $k$ is the total number of creative categories, $\sigma_i$ is the corresponding learnable weight, and $Loss_i$ is the loss of the i-th 
predicted creative element. In this way, the loss of multiple creative elements can be adaptively adjusted.

\section{EXPERIMENTS}

We conduct several groups of experiments to verify the model 
proposed, with the purpose of answering the following research
questions:

\textbf{RQ1} Does our proposed method outperform the baseline methods 
in the creative element selection problem?

\textbf{RQ2} How do the two modules in the model work for modeling the relationship between creative elements?

\textbf{RQ3} How does the model perform when deployed online?

\subsection{Datasets and Evaluation}

\subsubsection{Datasets.} Since there doesn’t exist an off-the-shelf benchmark dataset for such creative selection tasks, we construct our dataset from real data online. Firstly, we set up a small amount of random traffic online, i.e., creative elements are randomly combined by type, so that we can obtain an unbiased copy of user preference data.

In our scenario, an ad is composed of a banner background image, a main title, and a sub-title, so the creative type is 3, i.e. $k=3$. Finally, we collect a total number of 7.4 million samples, containing 5,494,110 users and 157,611 shops. We use 90\% of the data for training and the rest for testing. And we truncate each type of the creative elements sequences length of each shop to 5 uniformly, 
that is, up to 125($5^3$) potential creative combinations are considered for each shop. 

\subsubsection{Evaluation Metrics.} For evaluation, traditional 
ranking evaluation metrics such as nDCG, MAP are not suitable, because this is a combinatorial modeling problem. 
We focus on the \textit{"best combination"} of a desired advertisement,
and the evaluation of creative combination needs to be considered in a 
holistic way.

\textbf{Hit Ratio (HR).} Hit ratio is a recall-based evaluation metric
that measures how much the real creative elements in an ad 
overlap with the predicted elements. The formula is shown as follows:
\begin{gather}
  HR= \frac{\sum_{i=1}^k w_i |C_p\cap C_g|}{\sum_{i=1}^k w_i |C_p\cup C_g|}
\end{gather}
where $k$ is the number of creative element types, 
$C_p$ and $C_g$ are the predicted result and the ground truth respectively, $w_i$ indicates different weights for creative 
elements.

\textbf{Precision (PR).} The concept of precision is defined as whether the actually clicked (positive sample) creative element is also included in the estimated ad creative containing $k$ creative elements, formulated as follows:
\begin{gather}
  PR=  \textstyle \sum_{i=1}^k w_i I(c_i\in C_g)
\end{gather}
where $k$ is the number of creative element types,
$w_i$ is the weight parameter for each type of 
creative element, $I(\cdot)\in \{0,1\}$ is the indicator function.

In our dataset, considering the visual influence of the three types of creatives, we set 
$w_1=0.5, w_2=0.3, w_3=0.2$, and the overall $HR$ and $PR$ are 
calculated through the average metric of all testing examples.

\subsubsection{Baselines.} We compare the proposed method(\textbf{CECS}) 
with several strategies commonly used in online creative selection scenarios (\textbf{Strategies} for short).
To evaluate the ability to tackle relationships between elements,
we regard the selection as a multi-task CTR prediction task (\textbf{Multi-CTR} for short).
Therefore, we combine CTR models like DNN\cite{chen2016deep}, DeepFM\cite{guo2017deepfm},
AutoInt\cite{song2019autoint} with multi-task learning mechanisms like shared-bottom,
MMoE\cite{ma2018modeling} and PLE\cite{tang2020progressive} to fully explore 
the offline effects.


\subsection{Performance Comparison (RQ1)}
Table \ref{tab:1} shows the performance of HR and PR for the dataset for 
different methods. 
In general, the CTR-based methods are more effective than the Strategies cause these models are better at generalizing features.
With the improvement of the Multi-CTR model, we see a 
corresponding increase in the HR and PR evaluation metrics. 
And the MMoE structure captures the relationships better than the shared-bottom and slightly better than the PLE structure.
Among them, the AutoInt+MMoE model achieves the best results because 
the self-attention mechanism in this model is applied to capture 
the correlations between the same type of creative elements. 
Nevertheless, our model still outperforms the baselines, indicating the interaction between creative elements can be better captured.

From the table, we can conclude that the interaction between 
the elements in the Multi-CTR methods are not sufficient.
This is because different creative elements are not displayed
 independently, but in a holistic way, and contributes to 
 the final visual effect of the advertisement.

\begin{table}
  \caption{The performance for different models.}
  \label{tab:1}
  \begin{tabular}{cccc}
    \toprule
    \multicolumn{2}{c}{Model} & HR & PR \\
    \midrule
    \multirow{3}{*}{Strategies} & E-greedy & 0.7886 & 0.6707\\
    ~& Popularity & 0.7912 & 0.6804\\
    ~& Preference & 0.8025 & 0.7141\\
    \hline
    \multirow{4}{*}{Multi-CTR} & DNN + shared-bottom & 0.8328 & 0.7669\\
    ~& DNN + MMoE & 0.8456 & 0.7783\\
    ~& DNN + PLE & 0.8402 & 0.7776\\
    ~& DeepFM + MMoE & 0.8491 & 0.7812\\
    ~& AutoInt + MMoE & 0.8537 & 0.7873 \\
    \hline
    \multicolumn{2}{c}{\textbf{CECS(ours.)}} &
    \textbf{0.8702}&\textbf{0.8028}\\
  \bottomrule
\end{tabular}
\end{table}


\subsection{Analysis of Model Designs (RQ2)}
Here, we focus on the attention and the cascade design in CEI and CES respectively. We conduct several ablation studies for the two designs, 
as is shown in Table \ref{tab:2}. The $\surd$ and $\times$ marks in the table indicate whether the above designs are used or not. When not 
using the CEI, we use MLP instead. Not using the CES means that no estimated element information is transmitted through the decoder process, so we omit the hidden vector $h_{i-1}$ from the previous CES.

It can be seen from the table that with the
plug-in of different designs, the HR and PR of the 
model increase. We see there's a higher improvement in the CES design, which mainly contributes to the performance. Because through the design, the relationships can be transferred to other candidates to make better selections. The improvement of the CES design means that more information lies among different types of creative elements, and affects users' preference when combined together.

\begin{table}
  \caption{Ablation experiments corresponding to different 
  model designs.}
  \label{tab:2}
  \begin{tabular}{cccc}
    \toprule
    CEI & CES & HR & PR\\
    \midrule
    $\times$ & $\times$ & 0.8473 & 0.7767\\
    $\surd$ & $\times$ & 0.8501 & 0.7826\\
    $\times$ & $\surd$ & 0.8669 & 0.7952\\
    $\surd$ & $\surd$ & \textbf{0.8702} & \textbf{0.8028}\\
  \bottomrule
\end{tabular}
\end{table}




\subsection{Online Results(RQ3)}
To evaluate how the model performs in real-world industrial scenarios, we conduct A/B tests online. 
The model online is an MMoE-based Multi-CTR model, it predicts different types of creatives at the same time. Our model focus on the relationships between creative elements, solving them in a cascade selection way. For the online time complexity problem, we shrink the feature pulling time before prediction, then the inference time is within an acceptable range. In the end, compared with the Multi-CTR method,  we further obtained an increase of \textbf{CTR+6.02\%} and \textbf{GMV+10.37\%}, which proved the effectiveness of the method in the real-world industrial scene.


\section{CONCLUSION}

In this paper, we investigate the creative element selection from the perspective of cascade creative selection, solving the problem using a holistic way. The cross-element interaction is applied, which captures the associations between different types of candidate creative elements.
And we proposed the cascade element selection to 
model the relationship between estimated elements. 
It was experimentally verified that the method outperforms 
the traditional strategies and Multi-task CTR prediction method. And the ablation study verified the importance of  
attention and cascading design in the framework. The model was deployed on real-world data and significantly improved CTR and GMV, which can be extended and applied in other advertising scenarios.

\bibliographystyle{ACM-Reference-Format}
\bibliography{sample-base}

\end{document}